\documentclass[preprint, 11pt, authoryear]{elsarticle}

\usepackage[T1]{fontenc}
\usepackage{microtype}
\usepackage{geometry}
\usepackage{booktabs}
\usepackage{tabularx}
\newcolumntype{Y}{>{\centering\arraybackslash}X}
\usepackage{makecell}
\newcommand*{\tabindent}{\hspace{3mm}}
\usepackage{caption}
\usepackage{bm}
\usepackage[hyphens]{url}
\usepackage[hidelinks]{hyperref}
\hypersetup{breaklinks=true}
\journal{Accident Analysis \& Prevention}

\begin{document}

\begin{frontmatter}

\title{Sugar and Stops in Drivers with Insulin-Dependent Type 1 Diabetes}

\author[inst1]{Ashirwad Barnwal\corref{cor1}}
\ead{ashirwad@iastate.edu}
\cortext[cor1]{Corresponding author}
\author[inst2]{Pranamesh Chakraborty}
\author[inst1]{Anuj Sharma}
\author[inst3]{Luis Riera-Garcia}
\author[inst1]{Koray Ozcan}
\author[inst1]{Sayedomidreza Davami}
\author[inst3]{Soumik Sarkar}
\author[inst4]{Matthew Rizzo}
\author[inst4]{Jennifer Merickel}

\affiliation[inst1]{
    organization = {Institute for Transportation (InTrans), Iowa State University},
    city = {Ames},
    postcode = {50010}, 
    state = {Iowa},
    country = {United States}
}
\affiliation[inst2]{
    organization = {Department of Civil Engineering, IIT Kanpur},
    city = {Kanpur},
    postcode = {208016}, 
    state = {Uttar Pradesh},
    country = {India}
}
\affiliation[inst3]{
    organization = {Department of Mechanical Engineering, Iowa State University},
    city = {Ames},
    postcode = {50011}, 
    state = {Iowa},
    country = {United States}
}
\affiliation[inst4]{
    organization = {Department of Neurological Sciences, University of Nebraska Medical Center},
    city = {Omaha},
    postcode = {68198}, 
    state = {Nebraska},
    country = {United States}
}

\begin{abstract}
\textbf{Background:} Diabetes is a major public health challenge, affecting millions of people worldwide. Abnormal physiology in diabetes, particularly hypoglycemia, can cause driver impairments that affect safe driving. While diabetes driver safety has been previously researched, few studies link real-time physiologic changes in drivers with diabetes to objective real-world driver safety, particularly at high-risk areas like intersections. To address this, we investigated the role of acute physiologic changes in drivers with type 1 diabetes mellitus (T1DM) on safe stopping at stop intersections. \textbf{Methods:} 18 T1DM drivers (21--52 years, \(\mu\) = 31.2 years) and 14 controls (21--55 years, \(\mu\) = 33.4 years) participated in a 4-week naturalistic driving study. At induction, each participant's personal vehicle was instrumented with a camera and sensor system to collect driving data (e.g., GPS, video, speed). Video was processed with computer vision algorithms detecting traffic elements (e.g., traffic signals, stop signs). Stop intersections were geolocated with clustering methods, state intersection databases, and manual review. Videos showing driver stop intersection approaches were extracted and manually reviewed to classify stopping behavior (full, rolling, and no stop) and intersection traffic characteristics. \textbf{Results:} Mixed-effects logistic regression models determined how diabetes driver stopping safety (safe vs. unsafe stop) was affected by 1) disease and 2) at-risk, acute physiology (hypo- and hyperglycemia). Diabetes drivers who were acutely hyperglycemic (\(\geq\)300 mg/dL) had 2.37 increased odds of unsafe stopping (95\% CI: 1.26--4.47, p = 0.008) compared to those with normal physiology. Acute hypoglycemia did not associate with unsafe stopping (p = 0.537), however the lower frequency of hypoglycemia (vs. hyperglycemia) warrants a larger sample of drivers to investigate this effect. Critically, presence of diabetes alone did not associate with unsafe stopping, underscoring the need to evaluate driver physiology in licensing guidelines. \textbf{Conclusion:} This study links acute, abnormal physiologic fluctuations in drivers with diabetes to driver safety based on unsafe stopping at stop-controlled intersections, providing recommendations for clinicians aimed at improving patient safety, fair licensing guidelines, and targets for developing advanced driver assistance systems.
\end{abstract}


\end{frontmatter}

\section{Background}
Diabetes has become a critical problem of public health, affecting over 10\% of the United States population and continuing to grow world-wide in prevalence \citep{centers_for_disease_control_and_prevention_diabetes_2015}. The World Health Organization reported that the number of adults with diabetes nearly quadrupled worldwide from 108 million to 422 million between 1980--2014 \citep{world_health_organization_global_2016}. These statistics are increasingly alarming in the context of driver safety given the elevated vehicle crash risk in drivers with diabetes compared to drivers without diabetes \citep{tregear_diabetes_2007}. Increased driving risk in diabetes is linked to hypoglycemia (low blood glucose, or sugar), especially in drivers with insulin-dependent type 1 diabetes mellitus (T1DM) \citep{cox_driving_2009, skurtveit_road_2009}. In T1DM, the incidence of hypoglycemia is often increased due to use of glucose-lowering medications (e.g., insulin) aimed at controlling hyperglycemia (high blood glucose) to reduce risk of diabetes-related co-morbidities (e.g., retinal, renal, peripheral nerve, and cerebrovascular dysfunction). Consequently, driving licensing authorities in many developed countries consider the use of glucose-lowering agents as a driver risk factor when assessing licensure eligibility \citep{graveling_driving_2015}.

Hypoglycemia (\(\leq\)70 mg/dL) is widely linked to multi-domain cognitive and psychomotor dysfunction (e.g., across attention, decision-making, spatial processing, motor, and other domains) that can persist for hours after blood glucose levels return to normal \citep{brands_effects_2005, frier_symptoms_2013, evans_delay_2000, mccrimmon_diabetes_2012, rizzo_impaired_2011, warren_hypoglycaemia_2005}. Severe hypoglycemia (blood glucose \(\leq\)54 mg/dL) can even lead to coma and requires immediate medical treatment \citep{american_diabetes_association_standards_2019}. In addition to hypoglycemia, severe hyperglycemia is also linked to cognitive, motor, and perceptual dysfunction \citep{cox_effects_2007, gonder-frederick_cognitive_2009, kovatchev_postprandial_2003, sommerfield_acute_2004}. These dysfunctions increase the risk of driver safety errors and vehicle crashes. Impaired self-awareness of impairment, which results from cognitive dysfunction and also blunting of the body's typical autonomic response to hypoglycemia in response to repeated hypoglycemic exposure \citep{cryer_mechanisms_2005}, may also play a role in diabetes crash risk by hindering the driver's ability to appropriately mitigate risk.

Most prior research investigating driver safety in diabetes has used controlled simulation \citep{cox_progressive_2000, stork_decision_2007}. While simulator data is reproducible and controlled, it may not translate well to real-world safety. Drivers often behave differently in a simulator due to observer effects, driving an unfamiliar car, lack of safety risk, and potentially being in an atypical physiologic state such as induced hypoglycemia. Other research has focused on retrospective crash record reviews that may be missing key information and lack information on the driver's acute physiology at the time of crash or citation \citep{tregear_diabetes_2007}.

This study advanced prior research by overcoming these limitations through real-world, contemporaneous, physiologic monitoring in T1DM drivers. We combined naturalistic driving data collected through sensor systems embedded in the driver's primary vehicle with the driver's own, real-time glucose data collected from wearable continuous glucose monitoring (CGM) devices to determine the impact of acute (in-vehicle) glucose fluctuations on unsafe stopping behaviors (stop sign violations like rolling or no stops) in T1DM drivers at stop-controlled intersections (hereafter referred to as stop intersections). We also included control participants to provide baseline data in the absence of T1DM. For the purposes of this paper, we defined unsafe behavior as a driver action that was in violation of traffic law (failure to fully stop at a stop sign).

Stop intersections are a high-risk geographic node, with almost 700,000 motor vehicle crashes each year in the U.S. and injuries in nearly one-third of the crashes \citep{national_highway_traffic_safety_administration_traffic_2017}. Research has linked crashes at stop intersections to unsafe stopping behavior. In particular, \cite{retting_analysis_2003} found that nearly 70\% of all motor vehicle crashes that occurred at stop intersections were due to unsafe stopping behavior. These errors can be linked to driver impairments, like cognitive and psychomotor dysfunction in T1DM \citep{national_highway_traffic_safety_administration_crash_2010}. These characteristics make stop intersections a key, safety node to investigate the role of disease and physiology on driver safety in diabetes.

\section{Hypothesis}
We tested the hypothesis that unsafe stopping behavior in T1DM drivers would be associated with 1) presence of T1DM disease and 2) acute, at-risk physiologic states (hypo- and hyperglycemia).

\section{Methods}

\subsection{Data sources}

\subsubsection{Participant eligibility}
36 participants (21--59 years, \(\mu\) = 33.3 years) with T1DM or without disease (controls) were recruited from Omaha, Nebraska and surrounding areas via community or clinic recruitment at the University of Nebraska Medical Center (UNMC). All participants (a) had a legal driver’s license, (b) had >3 years of driving experience, (c) drove for at least 1 hour or 50 miles per week, (d) drove a single car \(\geq\)90\% of the driving time, and (e) had a car insurance for the study vehicle. Control participants without diabetes provided baseline data on driver behavior in the absence of diabetes. Control drivers were recruited to have similar demographic and health characteristics as T1DM participants without presence of diabetes. T1DM and control participants were matched for age (within 6 years), gender, education (within 2 years), and driving season (winter vs. not winter). Participants consented to study participation in accordance with UNMC's Institutional Review Board (UNMC IRB \# 462-16-FB).

Study eligibility was assessed at induction through medical history, physical examination (administered by a trained endocrinologist), and blood labs panel (Hemoglobin A1c [HbA1c], Glomerular Filtration Rate [GFR], and Thyroid-Stimulating Hormone [TSH]). T1DM participants had a type 1 diabetes diagnosis at induction, took insulin daily, self-reported hypoglycemia at least twice weekly, and had an HbA1c <12\% \citep{american_diabetes_association_standards_2019}. Control participants had no diagnosis history of diabetes and had an HbA1c <6\%. Medical conditions (peripheral nerve, eye, renal, neurological, and major psychiatric diseases) or concurrent medication usage (narcotics, sedating antihistamines, and major psychoactive medication) that could independently or co-morbidly significantly confound driver behavior or diabetes risk were excluded for all participants. All participants had safe vision for driving (near or far, uncorrected or corrected visual acuity <20/50 OU) per Nebraska licensure regulations. Four participants were excluded: 2 control participants due to presence of undiagnosed diabetes based on HbA1c tests and 2 T1DM participants due to vehicle incompatibility with the study's driving systems. The final sample was 32 participants (T1DM: \textit{N} = 18; Control: \textit{N} = 14).

\subsubsection{Study design}
\textbf{Driving:} Each participant participated for a 4-week period. Each participant's personal vehicle was instrumented with a ``Black Box'' instrumentation system to collect data on their typical driving habits. Participants were instructed to drive as they typically would. The system contained (a) two cameras, one positioned inward (facing the driver) and the other facing outward to capture the forward roadway; (b) a microphone inside the vehicle; (c) a Global Positioning System (GPS) receiver to record location, time, heading, and speed; (d) accelerometer sensors to capture vehicle's longitudinal, lateral, and vertical accelerations; and (e) an On-Board Diagnostics system port to record data on vehicle speed, throttle, and intake. Driving data were collected at one-second intervals (1 Hz) from on- to off-ignition.

\textbf{In-Laboratory Assessments:} Demographic (age, gender, race/ethnicity, marital status, socioeconomic status) and visual function (contrast sensitivity [Early Treatment Diabetic Retinopathy Study (ETDRS), 2.5\% OU] and near/far visual acuity [ETDRS OU]) data were collected at study start, in addition to medical history and medication usage. Table \ref{tab:part-smry-table} provides a summary of study participant's primary demographic and medical characteristics.

\begin{table}[!htbp]
\centering
\small
\caption{Summary of participant demographic and medical characteristics}
\label{tab:part-smry-table}
\begin{tabular}{@{}lccc@{}}
\toprule
 & \textbf{T1DM (\textit{N} = 18)} & \textbf{Control (\textit{N} = 14)} & \textbf{Total (\textit{N} = 32)} \\ \midrule
\textbf{Age (years)} &  &  &  \\
\tabindent Mean (SD) & 31.22 (9.67) & 33.36 (10.45) & 32.16 (9.91) \\
\tabindent Range & 21.00--52.00 & 21.00--55.00 & 21.00--55.00 \\
\textbf{Gender} &  &  &  \\
\tabindent Female & 11 (61.1\%) & 10 (71.4\%) & 21 (65.6\%) \\
\tabindent Male & 7 (38.9\%) & 4 (28.6\%) & 11 (34.4\%) \\
\textbf{Race} &  &  &  \\
\tabindent Asian & 0 (0.0\%) & 1 (7.1\%) & 1 (3.1\%) \\
\tabindent White & 18 (100.0\%) & 13 (92.9\%) & 31 (96.9\%) \\
\textbf{Driving experience (years)} &  &  &  \\
\tabindent Mean (SD) & 15.61 (9.59) & 16.07 (9.86) & 15.81 (9.55) \\
\tabindent Range & 6.00--36.00 & 4.00--33.00 & 4.00--36.00 \\
\textbf{HbA1c (\%)} &  &  &  \\
\tabindent Mean (SD) & 7.79 (1.05) & 5.19 (0.30) & 6.65 (1.54) \\
\tabindent Range & 6.80--11.30 & 4.70--5.70 & 4.70--11.30 \\ \bottomrule
\end{tabular}
\end{table}

\textbf{CGM:} T1DM participants were provided and trained in the use of a Food and Drug Administration (FDA) approved Dexcom G4 PLATINUM Professional Continuous Glucose Monitor (CGM) to wear for the 4-week study period. The Dexcom CGM is a wearable device that was situated on the abdomen of each T1DM participant to continuously measure glucose levels underneath the skin every 5 minutes. CGMs provide data on overall glucose level, rate of change, and direction of change. The CGM monitored only and did not provide any feedback to participants on glucose levels.

\subsubsection{CGM data processing}
The quality of data collected through Dexcom CGMs and participant compliance with device use were determined per FDA recommendation \citep{center_for_devices_and_radiological_health_fda_2014}. T1DM participants were compliant with CGM use, as evident from a) the low percentages of missing glucose data per participant (\(\mu\) = 5.5\%; range: 1.7\%--10.2\%) which is well within FDA guidelines requiring <25\% of missing CGM data for analysis and b) participants' compliance with CGM calibration procedures that required twice daily glucometer-sampled glucose levels to be entered into the CGM for device calibration (participants calibrated, on average, 2.1 times daily [range: 0--9 times]). For data quality, CGM data was post-processed to remove physiologically impossible glucose levels such as those that exceeded a rate of change of over 25\% within a 15-min time span, per FDA guidelines \citep{center_for_devices_and_radiological_health_fda_2014}. Post-processing quality checks resulted in, on average, 2.1\% of CGM data removed per participant. CGM data was merged with driving data by up-sampling CGM data to match the frequency of driving data (from 5 min to 1 Hz) and then merging the two data sets by timestamp. After merging CGM and driving data, 14.9\% of drives (\textit{N} = 271) were discarded due to missing glucose data.

\subsubsection{Glucose variables used for modeling}
The driver's \textbf{acute (in-vehicle) glucose state} was determined from CGM data and categorized as a) \textbf{hypoglycemic} (\(\leq\)70 mg/dL, \textit{N} = 65), b) \textbf{normal} (71--179 mg/dL, \textit{N} = 1098), or c) severely \textbf{hyperglycemic} (\(\geq\)300 mg/dL, \textit{N} = 381 [hereafter hyperglycemic]) following the American Diabetes Association (ADA) standards of medical care guidelines \citep{american_diabetes_association_6_2021}. We considered hypo- and hyperglycemic states as ``at-risk'' physiologic states.

\subsection{Stop intersection identification}
Stop intersections were identified within Nebraska and Iowa, the primary geographic driving locations for this study and containing 95.5\% of drives collected. To identify stop intersections visited by participants for modeling, we used a 2-step approach: (1) for intersections in Nebraska, we developed a data-driven, clustering based approach because Nebraska does not maintain an intersection database and (2) for intersections in Iowa, we used the intersection database for Iowa maintained at the Institute for Transportation at Iowa State University to geolocate driver-visited Iowa stop intersections.

\begin{figure}[!htbp]
\centering
\includegraphics{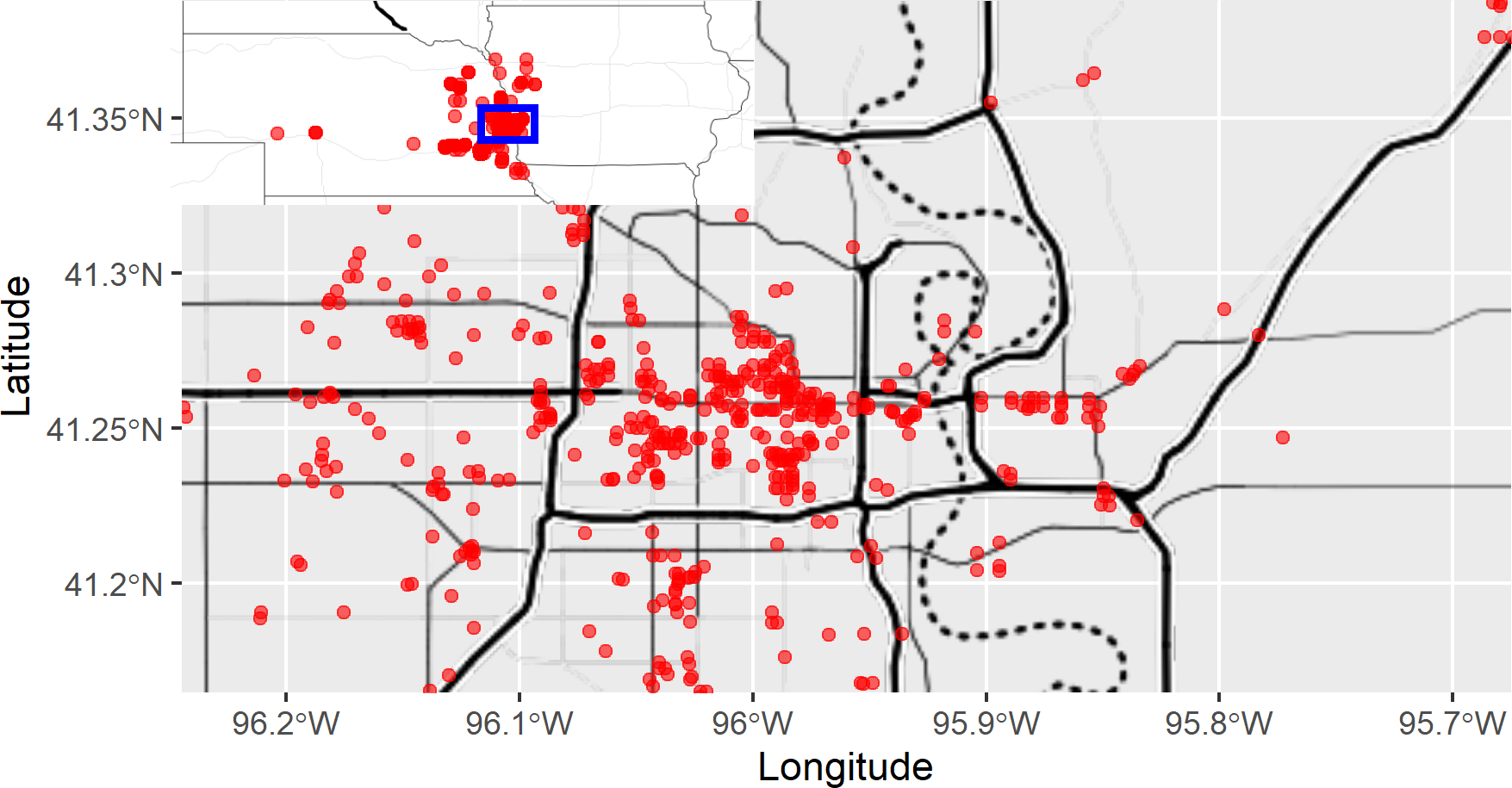}
\caption{Geographic locations of validated stop intersections visited by study participants}
\label{fig:stop-intxn-map}
\end{figure}

Nebraska stop intersections were identified by using a computer vision model capable of detecting and localizing 42 traffic element classes (across class mean average precision [mAP] = 79.8\%). Class examples included stop signs, traffic signs and lights, pedestrians, vehicles, buses, and bicycles. The model utilized a deep learning architecture for computer vision called Fast Region-based Convolutional Network [Fast R-CNN] \citep{girshick_fast_2015} and was trained on Black Box forward roadway video. The computer vision data was merged with the Black Box data using timestamps to get the geographic locations of stop sign detections. The resulting tabular data was then used to create a geographical representation of participant stop sign encounters (hereafter referred to as point features) using the ``sf'' R package [version 0.9.6; \cite{pebesma_simple_2018}]. Next, extracted point features were clustered using the Density-Based Spatial Clustering of Applications with Noise (DBSCAN) method implemented within the ``dbscan'' R package [version 1.1.5; \cite{hahsler_dbscan_2019}]. Cluster centers were calculated using Weizfeld's algorithm \citep{vardi_multivariate_2000} implemented within the ``Gmedian'' R package [version 1.2.5; \cite{cardot_gmedian_2020}] to geolocate stop intersections in Nebraska. The output of this process was a list of probable stop intersections in Nebraska (cluster detections: \textit{N} = 382; noise detections: \textit{N} = 681). Manual review using Google Earth was used to verify stop intersections and to classify intersection type (all-way or minor-road-only). The final, validated stop intersection list from Nebraska was merged with stop intersections identified from Iowa intersection databases (\textit{N} = 96). A total of 708 stop intersections [all-way (\textit{N} = 162), minor-road-only (\textit{N} = 546)] were visited by the participants and their geographic distribution is highlighted in Figure \ref{fig:stop-intxn-map}. The blue box represents the geographic areas (primarily Omaha, Nebraska and Council Bluffs, Iowa) that had the highest density of driver-visited stop intersections.

Black Box forward roadway videos showing each drivers' approach to the verified stop intersections was extracted. Extracted video clips contained each driver's stop intersection approach from 300 ft upstream and 200 ft downstream (based on Black Box GPS data). Clips were further manually reviewed to classify the stopping behavior as full stop (stopped vehicle at stop sign for \(\geq\)2 seconds), rolling stop (reduced vehicle speed at stop sign), and no stop (no change in vehicle speed at stop sign). Traffic variables that could confound driver behavior were also coded, including the presence or absence of a lead vehicle, a crossing vehicle, or a crossing pedestrian. For each traffic variable, whether the driver altered their behavior (yes/no; defined as a change in vehicle speed due to another vehicle or pedestrian) was also coded and was denoted as “\textbf{with effect}” and “\textbf{without effect}” for yes/no answers, respectively. The driver's identity as the consented research participant was also verified by inspecting Black Box cabin video corresponding to the stop sign encounter. Traffic and driver identity variables were defined for data selection purposes to exclude confounded or no-driver data and were not used in analysis. A descriptive summary of the forward roadway video variables at each stop intersection encounter is provided in Table \ref{tab:video-vars-smry-table}.

\begin{table}[!htbp]
\centering
\small
\caption{Summary of variables for data selection extracted through manual review. “Effect” denotes whether or not the participant altered their vehicle speed due to the presence of this variable.}
\label{tab:video-vars-smry-table}
\begin{tabular}{@{}lccc@{}}
\toprule
 & \textbf{\makecell{Safe (Full Stop) \\ (\textit{N} = 1140)}} & \textbf{\makecell{Unsafe \\ (80.8\% Rolling, 19.2\% No Stop) \\ (\textit{N} = 1864)}} & \textbf{\makecell{Total \\ (\textit{N} = 3004)}} \\ \midrule
\textbf{\makecell[tl]{Lead vehicle \\ status}} &  &  &  \\
\tabindent None present & 861 (75.5\%) & 1500 (80.5\%) & 2361 (78.6\%) \\
\tabindent Present with effect & 214 (18.8\%) & 240 (12.9\%) & 454 (15.1\%) \\
\tabindent Present without effect & 65 (5.7\%) & 124 (6.7\%) & 189 (6.3\%) \\
\textbf{\makecell[tl]{Crossing vehicle \\ status}} &  &  &  \\
\tabindent None present & 336 (29.5\%) & 1317 (70.7\%) & 1653 (55.0\%) \\
\tabindent Present with effect & 681 (59.7\%) & 117 (6.3\%) & 798 (26.6\%) \\
\tabindent Present without effect & 123 (10.8\%) & 430 (23.1\%) & 553 (18.4\%) \\
\textbf{\makecell[tl]{Crossing pedestrian \\ status}} &  &  &  \\
\tabindent None present & 1102 (96.7\%) & 1838 (98.6\%) & 2940 (97.9\%) \\
\tabindent Present with effect & 16 (1.4\%) & 7 (0.4\%) & 23 (0.8\%) \\
\tabindent Present without effect & 22 (1.9\%) & 19 (1.0\%) & 41 (1.4\%) \\
\textbf{\makecell[tl]{Is primary participant \\ driving?}} &  &  &  \\
\tabindent No & 96 (8.4\%) & 241 (12.9\%) & 337 (11.2\%) \\
\tabindent Yes & 1044 (91.6\%) & 1623 (87.1\%) & 2667 (88.8\%) \\ \bottomrule
\end{tabular}
\end{table}

Prior to modeling, data selection variables (Table \ref{tab:video-vars-smry-table}) were used to select a subset of data for modeling in which 1) the driver was consented research participant (88.8\% of stop intersection encounters) or 2) driver stop sign encounters in which their behavior was not altered due to a confounding vehicle or pedestrian (62.2\%) to ensure that valid comparisons could be made across stop sign encounters for modeling. The final dataset included 1,660 stop sign encounters.

\subsubsection{Stop variables used for modeling}
After data selection, final stopping behavior was categorized as either safe (\textit{N} = 347) or unsafe (\textit{N} = 1313) for modeling. Unsafe stops included rolling and no stops. These categories were combined to improve power due to the low frequency of ``no stop'' intersection encounters.

\section{Model structure}

\subsection{Modeling overview}
Models assessed 1) \textbf{disease status:} the role of disease on stopping behavior (comparing the stopping behavior of a T1DM driver to a control driver) and 2) \textbf{acute physiology:} whether a T1DM participant was more or less likely to make a safe or unsafe stop when they were acutely hypo- or hyperglycemic as compared to a) when they were in a normal glucose state or b) a control driver. Each participant had multiple stop sign encounters (\(\mu\) = 53.5; range: 9--127). Mixed-effects logistic regression (MELR) models were used in R's \citep{r_core_team_r_2021} ``lme4'' package [version 1.1.25; \cite{bates_fitting_2015}] to model stopping responses \(Y_{ij}\,(1 = \mathrm{unsafe}, \, 0 = \mathrm{safe})\) of participant \(i\,(i = 1, \ldots, N)\) for stop sign encounters \(j\,(j = 1, \ldots, n_{i})\) as a function of \(p\) covariates. The MELR equation is given as follows:

\[
\log\left[\frac{\Pr\left(Y_{ij} = 1\right)}{1 - \Pr\left(Y_{ij} = 1\right)}\right] = \bm{x}_{\bm{ij}}^{\prime}\bm{\beta}+v_{i} 
\]

where:

\begin{itemize}
\item
  \(\bm{x}_{\bm{ij}}\) is a column vector of \((p + 1)\) covariates,
\item
  \(\bm{\beta}\) is a column vector of \((p + 1)\) regression coefficients,
\item
  \(v_{i}\) is the participant random effects term assumed to be distributed as \(\mathcal{NID}\left(0, \sigma_{v}^{2}\right)\)
\end{itemize}

The disease status and the acute physiology models included covariates defining the disease status of the driver (T1DM vs. control) and the acute glycemic episode (control, normal, hypo, hyper). No other covariates were used in modeling.

\subsection{Model building procedure}
After deciding on the statistical modeling approach, we followed the following sequence of steps to build each model:

\begin{enumerate}
\item
  \textbf{Data partitioning:} Data was partitioned into subsets for modeling.
\item
  \textbf{Random effects assessment:} MELR models with and without intersection random effects (accounting for intersection-level variation) were built. All models included by-participant random effects (accounting for participant-level variation). Models were selected using analysis of variance (ANOVA) model comparison tests to determine optimal random effects structure (by-participant vs. by-participant and by-intersection random intercepts).
\item
  \textbf{Outlier data assessment:} Next, we assessed models for undue outlier influence. After potential outlier identification, model results with and without data from potential outlier were compared.
\end{enumerate}

\subsubsection{Outlier data assessment}
To determine data outliers that may obscure model results, we computed Cook's distance \citep{cook_influential_1979, cook_detection_1977} after iteratively refitting models across participants and intersections (grouping variables). Outlier data was discarded on (a) visual inspection of Cook's distances and (b) a thresholded Cook's distance of >0.5 \citep{pardoe_applied_2012}.

\subsection{Model 1: Disease status}
We first assessed the role of T1DM disease on unsafe stopping behavior.

\subsubsection{Data partitioning}
Data was partitioned into 2 sets: (1) \textbf{DM-All:} all T1DM and control data and (b) \textbf{DM-Norm:} T1DM data from T1DM drivers who were in acutely normal physiologic states and all control data. Descriptive statistics for DM-All and DM-Norm data sets are presented in Table \ref{tab:disease-descr-table}.

\begin{table}[!htbp]
\small
\centering
\caption{Overview of data used in disease status models}
\label{tab:disease-descr-table}
\begin{tabularx}{\textwidth}{lc * {5}{Y}}
\toprule
 & \multicolumn{3}{c}{\textbf{\makecell{DM-All \\ {[Control + T1DM (All Episodes)]}}}} & \multicolumn{3}{c}{\textbf{\makecell{DM-Norm \\ {[Control + T1DM (Normal Episodes)]}}}} \\
 \cmidrule(lr){2-4} \cmidrule(l){5-7}
 & \textbf{Total} & \textbf{Safe} & \textbf{Unsafe} & \textbf{Total} & \textbf{Safe} & \textbf{Unsafe} \\
 & \textbf{\textit{N} = 1660} & \textbf{\textit{N} = 347} & \textbf{\textit{N} = 1313} & \textbf{\textit{N} = 1167} & \textbf{\textit{N} = 260} & \textbf{\textit{N} = 907} \\ \midrule
\makecell[tl]{\textbf{Participant} \\ \textbf{type, N (\%)}} &  &  &  &  &  &  \\
\tabindent T1DM & \makecell{1054 \\ (63.5\%)} & \makecell{201 \\ (57.9\%)} & \makecell{853 \\ (65.0\%)} & \makecell{561 \\ (48.1\%)} & \makecell{114 \\ (43.8\%)} & \makecell{447 \\ (49.3\%)} \\
\tabindent Control & \makecell{606 \\ (36.5\%)} & \makecell{146 \\ (42.1\%)} & \makecell{460 \\ (35.0\%)} & \makecell{606 \\ (51.9\%)} & \makecell{146 \\ (56.2\%)} & \makecell{460 \\ (50.7\%)} \\ \bottomrule
\end{tabularx}
\end{table}

The data in Table \ref{tab:disease-descr-table} provides a high-level overview of how the distribution of unsafe stopping responses changes across participant types with and without data from T1DM participants under acutely abnormal physiologic states.

\subsubsection{Random effect assessment}
For each data partition (DM-All, DM-Norm), MELR models with (\textbf{DM-All1, DM-Norm1}) and without by-intersection random intercepts (\textbf{DM-All0, DM-Norm0}) were fit (Table \ref{tab:disease-prelim-models}). All models included by-participant random intercepts. Based on ANOVA model comparison, the models with by-intersections random intercepts (DM-All1 \(\left[\chi\left(1\right) = 27.09, \, p = \ <.001\right]\); DM-Norm1 \(\left[\chi\left(1\right) = 12.58, \, p = \ <.001\right]\)) provided better fit, so these models were considered further for outlier data assessment.

\begin{table}[!htbp]
\setlength{\tabcolsep}{0.5pt}
\fontsize{9.5}{11}\selectfont
\centering
\caption{Disease status model outputs comparing random effect structures}
\label{tab:disease-prelim-models}
\begin{tabularx}{\textwidth}{lc * {7}{Y}} 
\toprule
 & \multicolumn{2}{c}{\textbf{\makecell{DM-All0 \\ (part ranef)}}} & \multicolumn{2}{c}{\textbf{\makecell{DM-All1 \\ (part + intxn ranef)}}} & \multicolumn{2}{c}{\textbf{\makecell{DM-Norm0 \\ (part ranef)}}} & \multicolumn{2}{c}{\textbf{\makecell{DM-Norm1 \\ (part + intxn ranef)}}} \\ 
\textit{Predictors} & \multicolumn{1}{c}{\textit{OR}} & \textit{p} & \multicolumn{1}{c}{\textit{OR}} & \textit{p} & \multicolumn{1}{c}{\textit{OR}} & \textit{p} & \multicolumn{1}{c}{\textit{OR}} & \textit{p} \\ \midrule
Intercept & \makecell{5.54 \\ (2.09--14.66)} & 0.001 & \makecell{6.70 \\ (2.31--19.45)} & <0.001 & \makecell{5.45 \\ (2.23--13.34)} & <0.001 & \makecell{6.45 \\ (2.45--16.97)} & <0.001 \\
\makecell[cl]{Participant \\ type: T1DM} & \makecell{1.31 \\ (0.38--4.55)} & 0.669 & \makecell{1.34 \\ (0.35--5.18)} & 0.673 & \makecell{0.94 \\ (0.30--3.02)} & 0.924 & \makecell{0.94 \\ (0.27--3.28)} & 0.929 \\
\multicolumn{2}{l}{\textbf{Random Effects}} & \multicolumn{7}{l}{} \\
\(\sigma^2\) & \multicolumn{2}{l}{3.29} & \multicolumn{2}{l}{3.29} & \multicolumn{2}{l}{3.29} & \multicolumn{2}{l}{3.29} \\
\(\tau_{00}\) & \multicolumn{2}{l}{2.60\textsubscript{part}} & \multicolumn{2}{l}{0.94\textsubscript{intxn}} & \multicolumn{2}{l}{2.15\textsubscript{part}} & \multicolumn{2}{l}{0.76\textsubscript{intxn}} \\
 & \multicolumn{2}{l}{} & \multicolumn{2}{l}{3.04\textsubscript{part}} & \multicolumn{2}{l}{} & \multicolumn{2}{l}{2.43\textsubscript{part}} \\
ICC & \multicolumn{2}{l}{0.44} & \multicolumn{2}{l}{0.55} & \multicolumn{2}{l}{0.40} & \multicolumn{2}{l}{0.49} \\
N & \multicolumn{2}{l}{31\textsubscript{part}} & \multicolumn{2}{l}{31\textsubscript{part}} & \multicolumn{2}{l}{30\textsubscript{part}} & \multicolumn{2}{l}{30\textsubscript{part}} \\
 & \multicolumn{2}{l}{} & \multicolumn{2}{l}{502\textsubscript{intxn}} & \multicolumn{2}{l}{} & \multicolumn{2}{l}{413\textsubscript{intxn}} \\
Observations & \multicolumn{2}{l}{1660} & \multicolumn{2}{l}{1660} & \multicolumn{2}{l}{1167} & \multicolumn{2}{l}{1167} \\
\makecell[tl]{Marginal R\textsuperscript{2} / \\ Conditional R\textsuperscript{2}} & \multicolumn{2}{l}{0.003 / 0.443} & \multicolumn{2}{l}{0.003 / 0.548} & \multicolumn{2}{l}{0.000 / 0.395} & \multicolumn{2}{l}{0.000 / 0.493} \\ \bottomrule
\end{tabularx}
\end{table}

The definition of different components of the random effects estimates shown in Table \ref{tab:disease-prelim-models} are as follows: (a) \(\sigma^2\) is the within-group (or residual) variance. For MELR models, its value is fixed to \(\pi^2/3\); (b) \(\tau_{00}\) is the between-group random intercept variance; (c) ICC is the intraclass correlation coefficient; and (d) N is the number of random effect groups. Additionally, “part”, “intxn”, and “ranef” are shorthands for participant, intersection, and random effects, respectively.

\subsubsection{Outlier data assessment}
\begin{figure}[!htbp]
\centering
\includegraphics{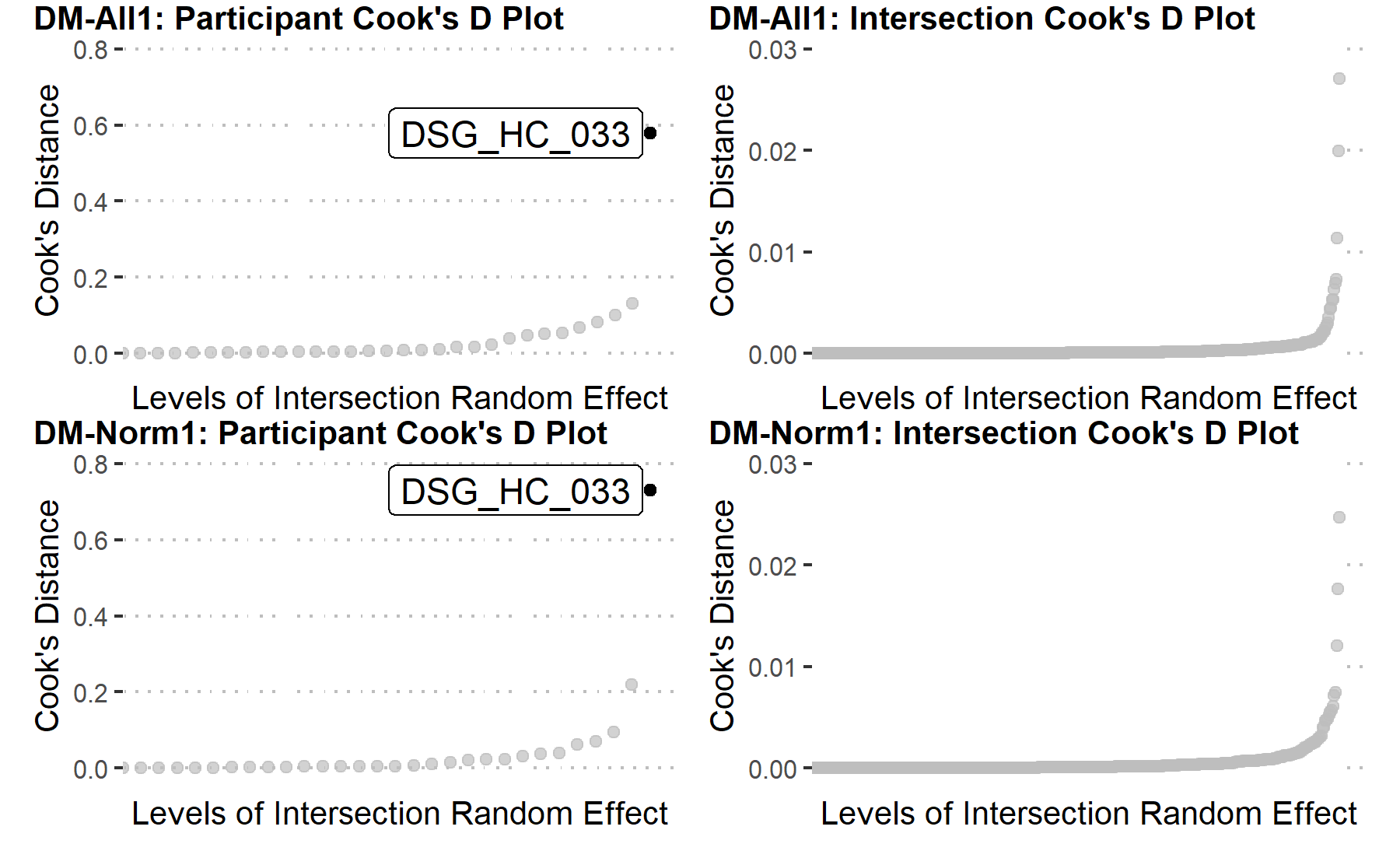}
\caption{Disease status outlier assessments using Cook’s distance}
\label{fig:disease-cooksd-plot}
\end{figure}

We computed and plotted (Figure \ref{fig:disease-cooksd-plot}) Cook's distance values for DM-All1 and DM-Norm1 models. Participant DSG\_HC\_033 was identified as a potential outlier participant in models following visual inspection and quantitative inspection (Cook's D = 0.58 [DM-All1] and Cook's D = 0.73 [DM-Norm1]). DM-All1 and DM-Norm1 models were re-run without data from the outlier participant (Table \ref{tab:disease-final-models}). While omitting the potential outlier participant changed the models odds ratio (variable Participant Type: T1DM), the p-values were not significantly impacted. Since model results did not significantly change as a result of the potential outlier participant, final model results included this participant.

\begin{table}[!htbp]
\setlength{\tabcolsep}{3.5pt}
\small
\centering
\caption{Comparison of disease model fits with and without the potential outlier participant}
\label{tab:disease-final-models}
\begin{tabular}{@{}llclclclc@{}}
\toprule
 & \multicolumn{2}{c}{\textbf{\makecell{DM-All1 \\ (preliminary)}}} & \multicolumn{2}{c}{\textbf{\makecell{DM-All1 \\ (infl part omitted)}}} & \multicolumn{2}{c}{\textbf{\makecell{DM-Norm1 \\ (preliminary)}}} & \multicolumn{2}{c}{\textbf{\makecell{DM-Norm1 \\ (infl part omitted)}}} \\
\textit{Predictors} & \multicolumn{1}{c}{\textit{OR}} & \textit{p} & \multicolumn{1}{c}{\textit{OR}} & \textit{p} & \multicolumn{1}{c}{\textit{OR}} & \textit{p} & \multicolumn{1}{c}{\textit{OR}} & \textit{p} \\ \midrule
Intercept & \multicolumn{1}{c}{\makecell{6.70 \\ (2.31--19.45)}} & <0.001 & \multicolumn{1}{c}{\makecell{10.96 \\ (4.33--27.72)}} & <0.001 & \multicolumn{1}{c}{\makecell{6.45 \\ (2.45--16.97)}} & <0.001 & \multicolumn{1}{c}{\makecell{10.51 \\ (4.66--23.70)}} & <0.001 \\
\makecell[cl]{Participant \\ type: T1DM} & \multicolumn{1}{c}{\makecell{1.34 \\ (0.35--5.18)}} & 0.673 & \multicolumn{1}{c}{\makecell{0.75 \\ (0.24--2.36)}} & 0.623 & \multicolumn{1}{c}{\makecell{0.94 \\ (0.27--3.28)}} & 0.929 & \multicolumn{1}{c}{\makecell{0.54 \\ (0.20--1.48)}} & 0.233 \\
\textbf{Random Effects} \\
\(\sigma^2\) & \multicolumn{2}{l}{3.29} & \multicolumn{2}{l}{3.29} & \multicolumn{2}{l}{3.29} & \multicolumn{2}{l}{3.29} \\
\(\tau_{00}\) & \multicolumn{2}{l}{0.94\textsubscript{intxn}} & \multicolumn{2}{l}{0.94\textsubscript{intxn}} & \multicolumn{2}{l}{0.76\textsubscript{intxn}} & \multicolumn{2}{l}{0.79\textsubscript{intxn}} \\
 & \multicolumn{2}{l}{3.04\textsubscript{part}} & \multicolumn{2}{l}{1.92\textsubscript{part}} & \multicolumn{2}{l}{2.43\textsubscript{part}} & \multicolumn{2}{l}{1.28\textsubscript{part}} \\
ICC & \multicolumn{2}{l}{0.55} & \multicolumn{2}{l}{0.46} & \multicolumn{2}{l}{0.49} & \multicolumn{2}{l}{0.39} \\
N & \multicolumn{2}{l}{31\textsubscript{part}} & \multicolumn{2}{l}{30\textsubscript{part}} & \multicolumn{2}{l}{30\textsubscript{part}} & \multicolumn{2}{l}{29\textsubscript{part}} \\
 & \multicolumn{2}{l}{502\textsubscript{intxn}} & \multicolumn{2}{l}{484\textsubscript{intxn}} & \multicolumn{2}{l}{413\textsubscript{intxn}} & \multicolumn{2}{l}{395\textsubscript{intxn}} \\
Observations & \multicolumn{2}{l}{1660} & \multicolumn{2}{l}{1571} & \multicolumn{2}{l}{1167} & \multicolumn{2}{l}{1078} \\
\makecell[tl]{Marginal R\textsuperscript{2} / \\  Conditional R\textsuperscript{2}} & \multicolumn{2}{l}{0.003 / 0.548} & \multicolumn{2}{l}{0.003 / 0.466} & \multicolumn{2}{l}{0.000 / 0.493} & \multicolumn{2}{l}{0.017 / 0.396} \\ \bottomrule
\end{tabular}
\end{table}

\subsubsection{Final disease status model results}
Final model results showed no significant influence from the potential outlier participant. Across all models, T1DM driver stopping behavior was not affected by disease, regardless of whether the behavior was compared to control drivers across all data or T1DM drivers in acutely, normal physiologic states.

\subsection{Model 2: Acute physiology}
To determine the role of acute, at-risk driver physiology on unsafe stopping behavior we assessed T1DM drivers in acutely hypo- and hyperglycemic states as compared to T1DM drivers in acutely, normal physiologic states and control drivers.

\subsubsection{Data partitioning}
Data was partitioned into 2 sets: (a) \textbf{All:} data from control and T1DM participants and (b) \textbf{DM:} data from just T1DM drivers (Table \ref{tab:acutephy-descr-table}).

\begin{table}[!htbp]
\centering
\small
\caption{Overview of data used in acute physiology models}
\label{tab:acutephy-descr-table}
\begin{tabularx}{\textwidth}{lc * {5}{Y}}
\toprule
 & \multicolumn{3}{c}{\textbf{\makecell{All \\ (Control + T1DM)}}} & \multicolumn{3}{c}{\textbf{\makecell{DM \\ (T1DM Only)}}} \\ 
 & \textbf{Total} & \textbf{Safe} & \textbf{Unsafe} & \textbf{Total} & \textbf{Safe} & \textbf{Unsafe} \\
 \cmidrule(lr){2-4} \cmidrule{5-7}
 & \textbf{\textit{N} = 1472} & \textbf{\textit{N} = 313} & \textbf{\textit{N} = 1159} & \textbf{\textit{N} = 866} & \textbf{\textit{N} = 167} & \textbf{\textit{N} = 699} \\ \midrule
\makecell[tl]{\textbf{Episode} \\ \textbf{type, N (\%)}} &  &  &  &  &  &  \\
\tabindent Control & \makecell{606 \\ (41.2\%)} & \makecell{146 \\ (46.6\%)} & \makecell{460 \\ (39.7\%)} & \makecell{0 \\ (0.00\%)} & \makecell{0 \\ (0.00\%)} & \makecell{0 \\ (0.00\%)} \\
\tabindent Hyper & \makecell{274 \\ (18.6\%)} & \makecell{47 \\ (15.0\%)} & \makecell{227 \\ (19.6\%)} & \makecell{274 \\ (31.6\%)} & \makecell{47 \\ (28.1\%)} & \makecell{227 \\ (32.5\%)} \\
\tabindent Hypo & \makecell{31 \\ (2.11\%)} & \makecell{6 \\ (1.92\%)} & \makecell{25 \\ (2.16\%)} & \makecell{31 \\ (3.58\%)} & \makecell{6 \\ (3.59\%)} & \makecell{25 \\ (3.58\%)} \\
\tabindent Normal & \makecell{561 \\ (38.1\%)} & \makecell{114 \\ (36.4\%)} & \makecell{447 \\ (38.6\%)} & \makecell{561 \\ (64.8\%)} & \makecell{114 \\ (68.3\%)} & \makecell{447 \\ (63.9\%)} \\ \bottomrule
\end{tabularx}
\end{table}

The data in Table \ref{tab:acutephy-descr-table} provides a high-level overview of how the distribution of unsafe stopping responses changes across different acute glycemic episodes with and without data from healthy controls.

\subsubsection{Random effect assessment}
ANOVA model comparisons were used to determine the optimal random effects structures for each dataset (Table \ref{tab:acutephy-prelim-models}). Models with by-intersection random intercepts (All1 [\(\chi\left(1\right) = 18.24, \, p = \ <.001\)]; DM1 [\(\chi\left(1\right) = 7.76, \, p = .005\)]) provided better fit to the data and were considered for outlier assessment.

\begin{table}[!htbp]
\setlength{\tabcolsep}{0.5pt}
\fontsize{9.5}{11}\selectfont
\centering
\caption{Acute physiology model outputs comparing random effect structures}
\label{tab:acutephy-prelim-models}
\begin{tabularx}{\textwidth}{lc * {7}{Y}}
\toprule
 & \multicolumn{2}{c}{\textbf{\makecell{All0 \\ (part ranef)}}} & \multicolumn{2}{c}{\textbf{\makecell{All1 \\ (part + intxn ranef)}}} & \multicolumn{2}{c}{\textbf{\makecell{DM0 \\ (part ranef)}}} & \multicolumn{2}{c}{\textbf{\makecell{DM1 \\ (part + intxn ranef)}}} \\
\textit{Predictors} & \multicolumn{1}{c}{\textit{OR}} & \textit{p} & \multicolumn{1}{c}{\textit{OR}} & \textit{p} & \multicolumn{1}{c}{\textit{OR}} & \textit{p} & \multicolumn{1}{c}{\textit{OR}} & \textit{p} \\ \midrule
Intercept & \multicolumn{1}{c}{\makecell{5.52 \\ (2.12–14.36)}} & <0.001 & \multicolumn{1}{c}{\makecell{6.62 \\ (2.34--18.78)}} & <0.001 & \multicolumn{1}{c}{\makecell{5.46 \\ (2.47--12.06)}} & <0.001 & \multicolumn{1}{c}{\makecell{6.18 \\ (2.65--14.43)}} & <0.001 \\
\makecell[cl]{Episode type: \\ Normal} & \multicolumn{1}{c}{\makecell{0.99 \\ (0.29--3.43)}} & 0.989 & \multicolumn{1}{c}{\makecell{0.97 \\ (0.25--3.70)}} & 0.963 & \multicolumn{1}{c}{} &  & \multicolumn{1}{c}{} &  \\
\makecell[cl]{Episode type: \\ Hypo} & \multicolumn{1}{c}{\makecell{0.68 \\ (0.14--3.29)}} & 0.630 & \multicolumn{1}{c}{\makecell{0.63 \\ (0.12--3.43)}} & 0.594 & \multicolumn{1}{c}{\makecell{0.68 \\ (0.24--1.94)}} & 0.475 & \multicolumn{1}{c}{\makecell{0.67 \\ (0.22--1.98)}} & 0.463 \\
\makecell[cl]{Episode type: \\ Hyper} & \multicolumn{1}{c}{\makecell{1.51 \\ (0.41--5.55)}} & 0.532 & \multicolumn{1}{c}{\makecell{1.49 \\ (0.37--6.03)}} & 0.578 & \multicolumn{1}{c}{\makecell{1.53 \\ (0.89--2.62)}} & 0.126 & \multicolumn{1}{c}{\makecell{1.54 \\ (0.88--2.68)}} & 0.130 \\
\multicolumn{2}{l}{\textbf{Random Effects}} & \multicolumn{7}{l}{} \\
\(\sigma^2\) & \multicolumn{2}{l}{3.29} & \multicolumn{2}{l}{3.29} & \multicolumn{2}{l}{3.29} & \multicolumn{2}{l}{3.29} \\
\(\tau_{00}\) & \multicolumn{2}{l}{2.49\textsubscript{part}} & \multicolumn{2}{l}{0.84\textsubscript{intxn}} & \multicolumn{2}{l}{2.45\textsubscript{part}} & \multicolumn{2}{l}{0.70\textsubscript{intxn}} \\
 & \multicolumn{2}{l}{} & \multicolumn{2}{l}{2.89\textsubscript{part}} & \multicolumn{2}{l}{} & \multicolumn{2}{l}{2.75\textsubscript{part}} \\
ICC & \multicolumn{2}{l}{0.43} & \multicolumn{2}{l}{0.53} & \multicolumn{2}{l}{0.43} & \multicolumn{2}{l}{0.51} \\
N & \multicolumn{2}{l}{30\textsubscript{part}} & \multicolumn{2}{l}{30\textsubscript{part}} & \multicolumn{2}{l}{18\textsubscript{part}} & \multicolumn{2}{l}{18\textsubscript{part}} \\
 & \multicolumn{2}{l}{} & \multicolumn{2}{l}{480\textsubscript{intxn}} & \multicolumn{2}{l}{} & \multicolumn{2}{l}{311\textsubscript{intxn}} \\
Observations & \multicolumn{2}{l}{1472} & \multicolumn{2}{l}{1472} & \multicolumn{2}{l}{866} & \multicolumn{2}{l}{866} \\
\makecell[tl]{Marginal R\textsuperscript{2} / \\ Conditional R\textsuperscript{2}} & \multicolumn{2}{l}{0.005 / 0.434} & \multicolumn{2}{l}{0.004 / 0.533} & \multicolumn{2}{l}{0.008 / 0.432} & \multicolumn{2}{l}{0.007 / 0.516} \\ \bottomrule
\end{tabularx}
\end{table}

\subsubsection{Outlier data assessment}
\begin{figure}[!htbp]
\centering
\includegraphics{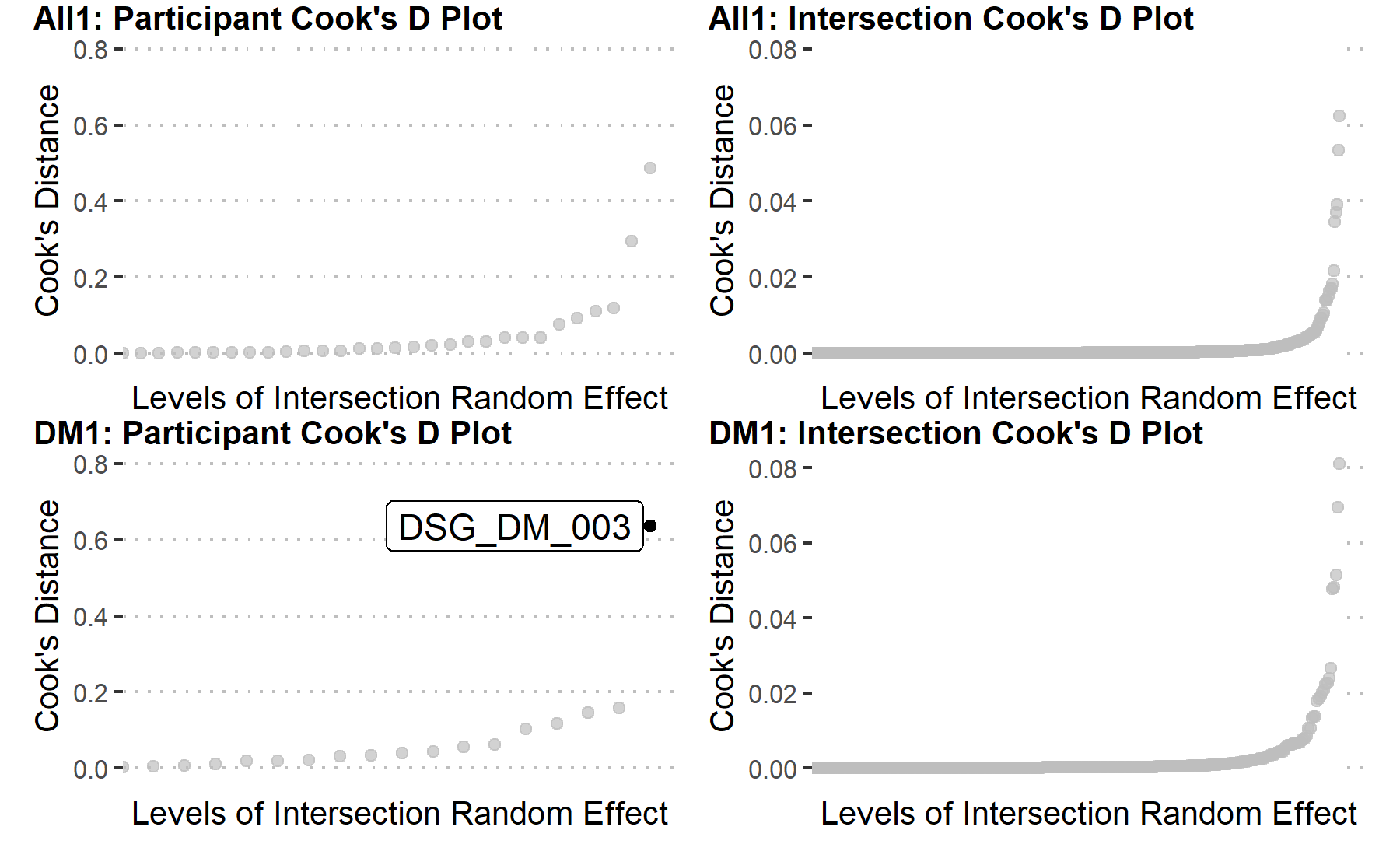}
\caption{Acute physiology outlier assessments using Cook’s distance}
\label{fig:acutephy-cooksd-plot}
\end{figure}

Cook's distance values for All1 and DM1 models were computed and plotted (Figure \ref{fig:acutephy-cooksd-plot}). Based on the visual and quantitative inspection, participant DSG\_DM\_003 was identified as a potential outlier only for the DM1 model (Cook's D = 0.64). So, we refitted DM1 model by omitting data from DSG\_DM\_003 participant (Table \ref{tab:acutephy-final-models}). While excluding this participant did not impact the effect of hypoglycemia on unsafe stopping, it significantly strengthened the effect of hyper episode (higher odds ratio, lower p-value), indicating that this participant disproportionately influenced model results. This participant was excluded from final models.

\begin{table}[!htbp]
\centering
\small
\caption{Comparison of acute physiology model results with and without the influential participant}
\label{tab:acutephy-final-models}
\begin{tabular}{@{}llclc}
\toprule
 & \multicolumn{2}{c}{\textbf{\makecell{DM1 \\ (preliminary)}}} & \multicolumn{2}{c}{\textbf{\makecell{DM1 \\ (infl part omitted)}}} \\
\textit{Predictors} & \multicolumn{1}{c}{\textit{OR}} & \textit{p} & \multicolumn{1}{c}{\textit{OR}} & \textit{p} \\ \midrule
Intercept & \multicolumn{1}{c}{\makecell{6.18 \\ (2.65--14.43)}} & <0.001 & \multicolumn{1}{c}{\makecell{6.56 \\ (3.06--14.06)}} & <0.001 \\
Episode type: Hypo & \multicolumn{1}{c}{\makecell{0.67 \\ (0.22--1.98)}} & 0.463 & \multicolumn{1}{c}{\makecell{0.71 \\ (0.24--2.09)}} & 0.537 \\
Episode type: Hyper & \multicolumn{1}{c}{\makecell{1.54 \\ (0.88--2.68)}} & 0.130 & \multicolumn{1}{c}{\makecell{2.37 \\ (1.26--4.47)}} & \textbf{0.008} \\
\textbf{Random Effects} & \multicolumn{2}{l}{} & \multicolumn{2}{l}{} \\
\(\sigma^2\) & \multicolumn{2}{l}{3.29} & \multicolumn{2}{l}{3.29} \\
\(\tau_{00}\) & \multicolumn{2}{l}{0.70\textsubscript{intxn}} & \multicolumn{2}{l}{0.56\textsubscript{intxn}} \\
 & \multicolumn{2}{l}{2.75\textsubscript{part}} & \multicolumn{2}{l}{1.98\textsubscript{part}} \\
ICC & \multicolumn{2}{l}{0.51} & \multicolumn{2}{l}{0.44} \\
N & \multicolumn{2}{l}{18\textsubscript{part}} & \multicolumn{2}{l}{17\textsubscript{part}} \\
 & \multicolumn{2}{l}{311\textsubscript{intxn}} & \multicolumn{2}{l}{292\textsubscript{intxn}} \\
Observations & \multicolumn{2}{l}{866} & \multicolumn{2}{l}{815} \\
Marginal R\textsuperscript{2} / Conditional R\textsuperscript{2} & \multicolumn{2}{l}{0.007 / 0.516} & \multicolumn{2}{l}{0.028 / 0.451} \\ \bottomrule
\end{tabular}
\end{table}

\subsubsection{Final acute physiology model results}
Final model results showed that T1DM drivers in acute hyperglycemic states are 2.37 times more likely (95\% CI: 1.26--4.47) to make unsafe stops than T1DM drivers in normal physiologic states. No significant differences in unsafe stopping behavior were found due to acute hypoglycemia.

\section{Conclusion}
This paper advances the need to link driver physiology to real-world driver safety behavior, in the context of diabetes and health broadly. While this was admittedly a pilot study and further research is merited, several promising findings have emerged. A key finding is that diabetes-presence alone, at least in the context of T1DM, was not associated with greater probabilities of unsafe stopping behavior. This finding has important implications. In the context of licensure evaluation, it suggests that licensing authorities should consider the individual's actual disease history above and beyond disease presence alone and that disease may not be a reliable predictor of an individual's on-road safety. This is in-line with the ADA's position statement that a diagnosis of diabetes is insufficient to evaluate a driver's abilities and that licensure evaluations relying broadly on diagnosis alone may inadvertently and unfairly restrict otherwise safe drivers \citep{american_diabetes_association_diabetes_2013}. In the context of clinical care, this stresses the need for doctors to consider a patient's disease history and individual characteristics when counseling and educating patients on driving with diabetes.

We find that acute hyperglycemia \(\geq\)300 mg/dL was strongly associated with unsafe stopping at stop intersections. In most prior driving research, hypoglycemia is more strongly associated with driver impairments and crash risk in diabetes than hyperglycemia. This finding suggests that hyperglycemia may produce greater impairments when an individual is operating amidst the complexities and challenges of the real-world than typically estimated from retrospective crash record review, self-report, and simulator studies \citep{cox_disruptive_2011, graveling_driving_2015}. Further research may explore if acute hyperglycemia at moderate levels (180--299 mg/dL) also affects unsafe stopping. Additionally, this study found that hypoglycemia was not associated with unsafe stopping, however, it is unclear if this result was driven by the relative lower frequency of hypoglycemia compared to hyperglycemia---limiting sample size in this pilot study. More research is needed to robustly determine the mechanisms that link acute physiologic changes in T1DM to driver safety, however we suggest that hyperglycemia, in addition to hypoglycemia, should be considered in diabetes driver evaluation and clinical safety assessments. Hyperglycemia also affects those with type 2 diabetes. Given the prevalence of type 2 diabetes \citep{centers_for_disease_control_and_prevention_national_2020} and its overwhelming public health burden, prospective research into real-world driver safety for type 2 diabetes is merited, along with consideration of a patient's degree of hyperglycemic control when assessing driver safety in diabetes.

Although our study shows some promising results, we highlight limitations with promise for future research. A primary limitation is our sample size of 32 drivers across 4-weeks of observation, particularly the low frequency of acute hypoglycemic episodes observed in this short observation period. Further research is needed in larger, more diverse populations, over longer periods to robustly determine how acute glucose abnormalities impact real-world driver safety. In context of driver physiology, prior literature suggests that driver impairment due to acute abnormal physiology may be mediated by effects from chronic exposure. Larger studies over longer durations can address how exposure to chronic, at-risk physiology mediates the risk of acute, at-risk physiology.

To expand this research effectively into larger populations of diabetes drivers and other diseases that impair driver physiology (e.g., sleep and cardiac dysfunctions, dementias, hypertension), automated big data processing methods to extract and label modeling variables (roadway features, driver behaviors) are needed. Even in this small pilot study, 33,550 driving miles were collected--limiting analytic ability due to labor-intensive, manual annotation needs. We suggest that computer vision, high performance computing, and cloud-based solutions offer promise.

To further contextualize driver safety, we suggest that further research investigate driver state and traffic environment covariates. Of key interest are surrounding traffic (e.g., a rolling stop in context of no intersection traffic vs. cross traffic) and the driver’s visual attention to the intersection (e.g., failing to make a full stop and not scanning surrounding traffic). These critical variables may interact with glucose-mediated impairments that affect a driver’s awareness of risk and cognitive (e.g., visuospatial, executive functioning) abilities. In addition, future research may also explore the role of environmental and roadway characteristics that may challenge impaired drivers and mediate safety such as weather, lighting, and road conditions. Our analyses focused on stopping behavior while limiting confounds like surrounding traffic to improve power and establish baseline links between glucose and driver safety in diabetes. Addition of these covariates will improve analytic ability to capture individual variation and illuminate the role of driver health and state factors that influence safety, in line with the promise of personalized medicine.

In summary, future research with larger and more diverse sample sizes is needed to further characterize driver safety across physiology and state. Particularly, we emphasize the need to account for driver state, traffic, and environmental variations in models to improve capture of individual health factors that influence real-world safety. Additional dependent measures (e.g., hard braking, distance from the stop sign when braking began) could also be analyzed to better understand the characteristics of the stopping behavior. Nevertheless, the results presented in this paper show promising methods for future research and demonstrate the need to capture real-time driver physiologic data when assessing behavior. As vehicle manufacturers continue to expand car capabilities, research of this line could guide development of advanced driver assistance systems aimed at supporting and keeping safe drivers with disease, ultimately improving health and quality of life in disease. Results also inform methods to inform clinical care and clinical trials with real-world patient function and safety data.

\section{Acknowledgements}
We gratefully acknowledge the Toyota Collaborative Safety Research Center for funding this study and the Mind \& Brain Health Labs at UNMC's Department of Neurological Sciences for leading study operations. In particular, we extend our thanks to Drs. Cyrus Desouza and Andjela Drincic for their invaluable clinical guidance on all aspects of this study, the Mind \& Brain Health Labs research staff, and Adam Hornig for recruiting and running study participants.

\Urlmuskip=0mu plus 1mu\relax
\bibliographystyle{elsarticle-harv}
\clearpage
\bibliography{references}

\begin{thebibliography}{38}
\expandafter\ifx\csname natexlab\endcsname\relax\def\natexlab#1{#1}\fi
\providecommand{\url}[1]{\texttt{#1}}
\providecommand{\href}[2]{#2}
\providecommand{\path}[1]{#1}
\providecommand{\DOIprefix}{doi:}
\providecommand{\ArXivprefix}{arXiv:}
\providecommand{\URLprefix}{URL: }
\providecommand{\Pubmedprefix}{pmid:}
\providecommand{\doi}[1]{\href{http://dx.doi.org/#1}{\path{#1}}}
\providecommand{\Pubmed}[1]{\href{pmid:#1}{\path{#1}}}
\providecommand{\bibinfo}[2]{#2}
\ifx\xfnm\relax \def\xfnm[#1]{\unskip,\space#1}\fi
\bibitem[{{American Diabetes
  Association}(2013)}]{american_diabetes_association_diabetes_2013}
\bibinfo{author}{{American Diabetes Association}}, \bibinfo{year}{2013}.
\newblock \bibinfo{title}{Diabetes and {Driving}}.
\newblock \bibinfo{journal}{Diabetes Care} \bibinfo{volume}{36},
  \bibinfo{pages}{S80--S85}.
\newblock \URLprefix
  \url{http://care.diabetesjournals.org/cgi/doi/10.2337/dc13-S080},
  \DOIprefix\doi{10.2337/dc13-S080}.
\bibitem[{{American Diabetes
  Association}(2019)}]{american_diabetes_association_standards_2019}
\bibinfo{author}{{American Diabetes Association}}, \bibinfo{year}{2019}.
\newblock \bibinfo{title}{Standards of {Medical} {Care} in {Diabetes}—2019
  {Abridged} for {Primary} {Care} {Providers}}.
\newblock \bibinfo{journal}{Clinical Diabetes} \bibinfo{volume}{37},
  \bibinfo{pages}{11--34}.
\newblock \URLprefix
  \url{http://clinical.diabetesjournals.org/lookup/doi/10.2337/cd18-0105},
  \DOIprefix\doi{10.2337/cd18-0105}.
\bibitem[{{American Diabetes
  Association}(2021)}]{american_diabetes_association_6_2021}
\bibinfo{author}{{American Diabetes Association}}, \bibinfo{year}{2021}.
\newblock \bibinfo{title}{6. {Glycemic} {Targets}: \textit{{Standards} of
  {Medical} {Care} in {Diabetes}—2021}}.
\newblock \bibinfo{journal}{Diabetes Care} \bibinfo{volume}{44},
  \bibinfo{pages}{S73--S84}.
\newblock \URLprefix
  \url{http://care.diabetesjournals.org/lookup/doi/10.2337/dc21-S006},
  \DOIprefix\doi{10.2337/dc21-S006}.
\bibitem[{Bates et~al.(2015)Bates, Mächler, Bolker and
  Walker}]{bates_fitting_2015}
\bibinfo{author}{Bates, D.}, \bibinfo{author}{Mächler, M.},
  \bibinfo{author}{Bolker, B.}, \bibinfo{author}{Walker, S.},
  \bibinfo{year}{2015}.
\newblock \bibinfo{title}{Fitting {Linear} {Mixed}-{Effects} {Models} {Using}
  \textbf{lme4}}.
\newblock \bibinfo{journal}{Journal of Statistical Software}
  \bibinfo{volume}{67}.
\newblock \URLprefix \url{http://www.jstatsoft.org/v67/i01/},
  \DOIprefix\doi{10.18637/jss.v067.i01}.
\bibitem[{Brands et~al.(2005)Brands, Biessels, de~Haan, Kappelle and
  Kessels}]{brands_effects_2005}
\bibinfo{author}{Brands, A.M.}, \bibinfo{author}{Biessels, G.J.},
  \bibinfo{author}{de~Haan, E.H.}, \bibinfo{author}{Kappelle, L.J.},
  \bibinfo{author}{Kessels, R.P.}, \bibinfo{year}{2005}.
\newblock \bibinfo{title}{The {Effects} of {Type} 1 {Diabetes} on {Cognitive}
  {Performance}: {A} meta-analysis}.
\newblock \bibinfo{journal}{Diabetes Care} \bibinfo{volume}{28},
  \bibinfo{pages}{726--735}.
\newblock \URLprefix
  \url{http://care.diabetesjournals.org/cgi/doi/10.2337/diacare.28.3.726},
  \DOIprefix\doi{10.2337/diacare.28.3.726}.
\bibitem[{Cardot(2020)}]{cardot_gmedian_2020}
\bibinfo{author}{Cardot, H.}, \bibinfo{year}{2020}.
\newblock \bibinfo{title}{Gmedian: {Geometric} {Median}, k-{Median}
  {Clustering} and {Robust} {Median} {PCA}}.
\newblock \URLprefix \url{https://CRAN.R-project.org/package=Gmedian}.
\bibitem[{{Center for Devices and Radiological
  Health}(2014)}]{center_for_devices_and_radiological_health_fda_2014}
\bibinfo{author}{{Center for Devices and Radiological Health}},
  \bibinfo{year}{2014}.
\newblock \bibinfo{title}{{FDA} {Summary} of safety and effectiveness data for
  {Dexcom} {G4} {PLATINUM} continuous glucose monitoring system}.
\newblock \bibinfo{type}{Technical Report} \bibinfo{number}{PMA P120005/S018}.
  U.S. Food and Drug Administration. \bibinfo{address}{White Oak, MD}.
\newblock \URLprefix
  \url{https://www.accessdata.fda.gov/cdrh_docs/pdf12/P120005S018b.pdf}.
\bibitem[{{Centers for Disease Control and
  Prevention}(2015)}]{centers_for_disease_control_and_prevention_diabetes_2015}
\bibinfo{author}{{Centers for Disease Control and Prevention}},
  \bibinfo{year}{2015}.
\newblock \bibinfo{title}{Diabetes {Report} {Card} 2014}.
\newblock \bibinfo{type}{Technical Report}. U.S. Department of Health and Human
  Services. \bibinfo{address}{Atlanta, GA}.
\newblock \URLprefix
  \url{https://www.cdc.gov/diabetes/pdfs/library/diabetesreportcard2014.pdf}.
\bibitem[{{Centers for Disease Control and
  Prevention}(2020)}]{centers_for_disease_control_and_prevention_national_2020}
\bibinfo{author}{{Centers for Disease Control and Prevention}},
  \bibinfo{year}{2020}.
\newblock \bibinfo{title}{National {Diabetes} {Statistics} {Report}}.
\newblock \bibinfo{type}{Technical Report}. U.S. Dept of Health and Human
  Services. \bibinfo{address}{Atlanta, GA}.
\bibitem[{Cook(1977)}]{cook_detection_1977}
\bibinfo{author}{Cook, R.D.}, \bibinfo{year}{1977}.
\newblock \bibinfo{title}{Detection of {Influential} {Observation} in {Linear}
  {Regression}}.
\newblock \bibinfo{journal}{Technometrics} \bibinfo{volume}{19},
  \bibinfo{pages}{15}.
\newblock \URLprefix
  \url{https://www.jstor.org/stable/1268249?origin=crossref},
  \DOIprefix\doi{10.2307/1268249}.
\bibitem[{Cook(1979)}]{cook_influential_1979}
\bibinfo{author}{Cook, R.D.}, \bibinfo{year}{1979}.
\newblock \bibinfo{title}{Influential {Observations} in {Linear} {Regression}}.
\newblock \bibinfo{journal}{Journal of the American Statistical Association}
  \bibinfo{volume}{74}, \bibinfo{pages}{169--174}.
\newblock \URLprefix
  \url{http://www.tandfonline.com/doi/abs/10.1080/01621459.1979.10481634},
  \DOIprefix\doi{10.1080/01621459.1979.10481634}.
\bibitem[{Cox et~al.(2009)Cox, Ford, Gonder-Frederick, Clarke, Mazze, Weinger
  and Ritterband}]{cox_driving_2009}
\bibinfo{author}{Cox, D.J.}, \bibinfo{author}{Ford, D.},
  \bibinfo{author}{Gonder-Frederick, L.}, \bibinfo{author}{Clarke, W.},
  \bibinfo{author}{Mazze, R.}, \bibinfo{author}{Weinger, K.},
  \bibinfo{author}{Ritterband, L.}, \bibinfo{year}{2009}.
\newblock \bibinfo{title}{Driving {Mishaps} {Among} {Individuals} {With} {Type}
  1 {Diabetes}: {A} prospective study}.
\newblock \bibinfo{journal}{Diabetes Care} \bibinfo{volume}{32},
  \bibinfo{pages}{2177--2180}.
\newblock \URLprefix
  \url{http://care.diabetesjournals.org/cgi/doi/10.2337/dc08-1510},
  \DOIprefix\doi{10.2337/dc08-1510}.
\bibitem[{Cox et~al.(2011)Cox, Ford, Ritterband, Singh and
  Gonder-Frederick}]{cox_disruptive_2011}
\bibinfo{author}{Cox, D.J.}, \bibinfo{author}{Ford, D.},
  \bibinfo{author}{Ritterband, L.}, \bibinfo{author}{Singh, H.},
  \bibinfo{author}{Gonder-Frederick, L.}, \bibinfo{year}{2011}.
\newblock \bibinfo{title}{Disruptive effects of hyperglycemia on driving in
  adults with type 1 \& 2 diabetes}, \bibinfo{publisher}{AMER DIABETES ASSOC
  1701 N BEAUREGARD ST, ALEXANDRIA, VA 22311-1717 USA}. pp.
  \bibinfo{pages}{A223--A223}.
\bibitem[{Cox et~al.(2000)Cox, Gonder-Frederick, Kovatchev, Julian and
  Clarke}]{cox_progressive_2000}
\bibinfo{author}{Cox, D.J.}, \bibinfo{author}{Gonder-Frederick, L.A.},
  \bibinfo{author}{Kovatchev, B.P.}, \bibinfo{author}{Julian, D.M.},
  \bibinfo{author}{Clarke, W.L.}, \bibinfo{year}{2000}.
\newblock \bibinfo{title}{Progressive hypoglycemia's impact on driving
  simulation performance. {Occurrence}, awareness and correction}.
\newblock \bibinfo{journal}{Diabetes Care} \bibinfo{volume}{23},
  \bibinfo{pages}{163--170}.
\newblock \URLprefix
  \url{http://care.diabetesjournals.org/cgi/doi/10.2337/diacare.23.2.163},
  \DOIprefix\doi{10.2337/diacare.23.2.163}.
\bibitem[{Cox et~al.(2007)Cox, McCall, Kovatchev, Sarwat, Ilag and
  Tan}]{cox_effects_2007}
\bibinfo{author}{Cox, D.J.}, \bibinfo{author}{McCall, A.},
  \bibinfo{author}{Kovatchev, B.}, \bibinfo{author}{Sarwat, S.},
  \bibinfo{author}{Ilag, L.L.}, \bibinfo{author}{Tan, M.H.},
  \bibinfo{year}{2007}.
\newblock \bibinfo{title}{Effects of {Blood} {Glucose} {Rate} of {Changes} on
  {Perceived} {Mood} and {Cognitive} {Symptoms} in {Insulin}-{Treated} {Type} 2
  {Diabetes}}.
\newblock \bibinfo{journal}{Diabetes Care} \bibinfo{volume}{30},
  \bibinfo{pages}{2001--2002}.
\newblock \URLprefix
  \url{https://diabetesjournals.org/care/article/30/8/2001/28490/Effects-of-Blood-Glucose-Rate-of-Changes-on},
  \DOIprefix\doi{10.2337/dc06-2480}.
\bibitem[{Cryer(2005)}]{cryer_mechanisms_2005}
\bibinfo{author}{Cryer, P.E.}, \bibinfo{year}{2005}.
\newblock \bibinfo{title}{Mechanisms of {Hypoglycemia}-{Associated} {Autonomic}
  {Failure} and {Its} {Component} {Syndromes} in {Diabetes}}.
\newblock \bibinfo{journal}{Diabetes} \bibinfo{volume}{54},
  \bibinfo{pages}{3592--3601}.
\newblock \URLprefix
  \url{http://diabetes.diabetesjournals.org/cgi/doi/10.2337/diabetes.54.12.3592},
  \DOIprefix\doi{10.2337/diabetes.54.12.3592}.
\bibitem[{Deary and Zammitt(2013)}]{frier_symptoms_2013}
\bibinfo{author}{Deary, I.J.}, \bibinfo{author}{Zammitt, N.N.},
  \bibinfo{year}{2013}.
\newblock \bibinfo{title}{Symptoms of {Hypoglycaemia} and {Effects} on {Mental}
  {Performance} and {Emotions}}, in: \bibinfo{editor}{Frier, B.M.},
  \bibinfo{editor}{Heller, S.R.}, \bibinfo{editor}{McCrimmon, R.J.} (Eds.),
  \bibinfo{booktitle}{Hypoglycaemia in {Clinical} {Diabetes}}.
  \bibinfo{publisher}{John Wiley \& Sons, Ltd}, \bibinfo{address}{Oxford, UK},
  pp. \bibinfo{pages}{23--45}.
\newblock \URLprefix \url{http://doi.wiley.com/10.1002/9781118695432.ch2},
  \DOIprefix\doi{10.1002/9781118695432.ch2}.
\bibitem[{Evans et~al.(2000)Evans, Pernet, Lomas, Jones and
  Amiel}]{evans_delay_2000}
\bibinfo{author}{Evans, M.L.}, \bibinfo{author}{Pernet, A.},
  \bibinfo{author}{Lomas, J.}, \bibinfo{author}{Jones, J.},
  \bibinfo{author}{Amiel, S.A.}, \bibinfo{year}{2000}.
\newblock \bibinfo{title}{Delay in onset of awareness of acute hypoglycemia and
  of restoration of cognitive performance during recovery}.
\newblock \bibinfo{journal}{Diabetes Care} \bibinfo{volume}{23},
  \bibinfo{pages}{893--897}.
\newblock \URLprefix
  \url{http://care.diabetesjournals.org/cgi/doi/10.2337/diacare.23.7.893},
  \DOIprefix\doi{10.2337/diacare.23.7.893}.
\bibitem[{Girshick(2015)}]{girshick_fast_2015}
\bibinfo{author}{Girshick, R.}, \bibinfo{year}{2015}.
\newblock \bibinfo{title}{Fast {R}-{CNN}}, in: \bibinfo{booktitle}{2015 {IEEE}
  {International} {Conference} on {Computer} {Vision} ({ICCV})},
  \bibinfo{publisher}{IEEE}, \bibinfo{address}{Santiago, Chile}. pp.
  \bibinfo{pages}{1440--1448}.
\newblock \URLprefix \url{http://ieeexplore.ieee.org/document/7410526/},
  \DOIprefix\doi{10.1109/ICCV.2015.169}.
\bibitem[{Gonder-Frederick et~al.(2009)Gonder-Frederick, Zrebiec, Bauchowitz,
  Ritterband, Magee, Cox and Clarke}]{gonder-frederick_cognitive_2009}
\bibinfo{author}{Gonder-Frederick, L.A.}, \bibinfo{author}{Zrebiec, J.F.},
  \bibinfo{author}{Bauchowitz, A.U.}, \bibinfo{author}{Ritterband, L.},
  \bibinfo{author}{Magee, J.C.}, \bibinfo{author}{Cox, D.J.},
  \bibinfo{author}{Clarke, W.L.}, \bibinfo{year}{2009}.
\newblock \bibinfo{title}{Cognitive {Function} {Is} {Disrupted} by {Both}
  {Hypo}- and {Hyperglycemia} in {School}-{AgedChildren} {With} {Type} 1
  {Diabetes}: {A} {Field} {Study}}.
\newblock \bibinfo{journal}{Diabetes Care} \bibinfo{volume}{32},
  \bibinfo{pages}{1001--1006}.
\newblock \URLprefix
  \url{https://diabetesjournals.org/care/article/32/6/1001/28678/Cognitive-Function-Is-Disrupted-by-Both-Hypo-and},
  \DOIprefix\doi{10.2337/dc08-1722}.
\bibitem[{Graveling and Frier(2015)}]{graveling_driving_2015}
\bibinfo{author}{Graveling, A.J.}, \bibinfo{author}{Frier, B.M.},
  \bibinfo{year}{2015}.
\newblock \bibinfo{title}{Driving and diabetes: problems, licensing
  restrictions and recommendations for safe driving}.
\newblock \bibinfo{journal}{Clinical Diabetes and Endocrinology}
  \bibinfo{volume}{1}, \bibinfo{pages}{8}.
\newblock \URLprefix
  \url{https://clindiabetesendo.biomedcentral.com/articles/10.1186/s40842-015-0007-3},
  \DOIprefix\doi{10.1186/s40842-015-0007-3}. \bibinfo{note}{tex.ids:
  gravelingDrivingDiabetesProblems2015a,
  gravelingDrivingDiabetesProblems2015b}.
\bibitem[{Hahsler et~al.(2019)Hahsler, Piekenbrock and
  Doran}]{hahsler_dbscan_2019}
\bibinfo{author}{Hahsler, M.}, \bibinfo{author}{Piekenbrock, M.},
  \bibinfo{author}{Doran, D.}, \bibinfo{year}{2019}.
\newblock \bibinfo{title}{\textbf{dbscan} : {Fast} {Density}-{Based}
  {Clustering} with \textit{{R}}}.
\newblock \bibinfo{journal}{Journal of Statistical Software}
  \bibinfo{volume}{91}.
\newblock \URLprefix \url{http://www.jstatsoft.org/v91/i01/},
  \DOIprefix\doi{10.18637/jss.v091.i01}.
\bibitem[{Kovatchev et~al.(2003)Kovatchev, Cox, Summers, Gonder-Frederick and
  Clarke}]{kovatchev_postprandial_2003}
\bibinfo{author}{Kovatchev, B.P.}, \bibinfo{author}{Cox, D.J.},
  \bibinfo{author}{Summers, K.H.}, \bibinfo{author}{Gonder-Frederick, L.A.},
  \bibinfo{author}{Clarke, W.L.}, \bibinfo{year}{2003}.
\newblock \bibinfo{title}{Postprandial glucose dynamics and associated symptoms
  in type 2 diabetes mellitus}.
\newblock \bibinfo{journal}{J Appl Res} \bibinfo{volume}{3},
  \bibinfo{pages}{449--458}.
\newblock \bibinfo{note}{Publisher: Citeseer}.
\bibitem[{McCrimmon et~al.(2012)McCrimmon, Ryan and
  Frier}]{mccrimmon_diabetes_2012}
\bibinfo{author}{McCrimmon, R.J.}, \bibinfo{author}{Ryan, C.M.},
  \bibinfo{author}{Frier, B.M.}, \bibinfo{year}{2012}.
\newblock \bibinfo{title}{Diabetes and cognitive dysfunction}.
\newblock \bibinfo{journal}{The Lancet} \bibinfo{volume}{379},
  \bibinfo{pages}{2291--2299}.
\newblock \URLprefix
  \url{https://linkinghub.elsevier.com/retrieve/pii/S0140673612603602},
  \DOIprefix\doi{10.1016/S0140-6736(12)60360-2}.
\bibitem[{{National Highway Traffic Safety
  Administration}(2010)}]{national_highway_traffic_safety_administration_crash_2010}
\bibinfo{author}{{National Highway Traffic Safety Administration}},
  \bibinfo{year}{2010}.
\newblock \bibinfo{title}{Crash {Factors} in {Intersection}-{Related}
  {Crashes}: {An} {On}-{Scene} {Perspective}}.
\newblock \bibinfo{type}{Technical Report} \bibinfo{number}{DOT HS 811 366}.
  U.S. Department of Transportation. \bibinfo{address}{Washington, DC}.
\newblock \URLprefix \url{https://trid.trb.org/View/1083638}.
  \bibinfo{note}{number: HS-811 366}.
\bibitem[{{National Highway Traffic Safety
  Administration}(2017)}]{national_highway_traffic_safety_administration_traffic_2017}
\bibinfo{author}{{National Highway Traffic Safety Administration}},
  \bibinfo{year}{2017}.
\newblock \bibinfo{title}{Traffic {Safety} {Facts} 2015: {A} {Compilation} of
  {Motor} {Vehicle} {Crash} {Data} from the {Fatality} {Analysis} {Reporting}
  {System} and the {General} {Estimates} {System}}.
\newblock \bibinfo{type}{Technical Report} \bibinfo{number}{DOT HS 812 384}.
  U.S. Department of Transportation. \bibinfo{address}{Washington, DC}.
\newblock \URLprefix \url{https://trid.trb.org/View/1465457}.
  \bibinfo{note}{number: DOT HS 812 384}.
\bibitem[{Pardoe(2012)}]{pardoe_applied_2012}
\bibinfo{author}{Pardoe, I.}, \bibinfo{year}{2012}.
\newblock \bibinfo{title}{Applied regression modeling}.
\newblock \bibinfo{edition}{2nd ed} ed., \bibinfo{publisher}{Wiley},
  \bibinfo{address}{Hoboken, NJ}.
\newblock \bibinfo{note}{Medium: electronic resource}.
\bibitem[{Pebesma(2018)}]{pebesma_simple_2018}
\bibinfo{author}{Pebesma, E.}, \bibinfo{year}{2018}.
\newblock \bibinfo{title}{Simple {Features} for {R}: {Standardized} {Support}
  for {Spatial} {Vector} {Data}}.
\newblock \bibinfo{journal}{The R Journal} \bibinfo{volume}{10},
  \bibinfo{pages}{439}.
\newblock \URLprefix
  \url{https://journal.r-project.org/archive/2018/RJ-2018-009/index.html},
  \DOIprefix\doi{10.32614/RJ-2018-009}.
\bibitem[{{R Core Team}(2021)}]{r_core_team_r_2021}
\bibinfo{author}{{R Core Team}}, \bibinfo{year}{2021}.
\newblock \bibinfo{title}{R: {A} {Language} and {Environment} for {Statistical}
  {Computing}}.
\newblock \bibinfo{publisher}{R Foundation for Statistical Computing},
  \bibinfo{address}{Vienna, Austria}.
\newblock \URLprefix \url{https://www.R-project.org/}. \bibinfo{note}{tex.ids=
  rcoreteamLanguageEnvironmentStatistical2020}.
\bibitem[{Retting et~al.(2003)Retting, Weinstein and
  Solomon}]{retting_analysis_2003}
\bibinfo{author}{Retting, R.A.}, \bibinfo{author}{Weinstein, H.B.},
  \bibinfo{author}{Solomon, M.G.}, \bibinfo{year}{2003}.
\newblock \bibinfo{title}{Analysis of motor-vehicle crashes at stop signs in
  four {U}.{S}. cities}.
\newblock \bibinfo{journal}{Journal of Safety Research} \bibinfo{volume}{34},
  \bibinfo{pages}{485--489}.
\newblock \URLprefix
  \url{https://linkinghub.elsevier.com/retrieve/pii/S0022437503000689},
  \DOIprefix\doi{10.1016/j.jsr.2003.05.001}.
\bibitem[{Rizzo(2011)}]{rizzo_impaired_2011}
\bibinfo{author}{Rizzo, M.}, \bibinfo{year}{2011}.
\newblock \bibinfo{title}{Impaired {Driving} {From} {Medical} {Conditions}: {A}
  70-{Year}-{Old} {Man} {Trying} to {Decide} if {He} {Should} {Continue}
  {Driving}}.
\newblock \bibinfo{journal}{JAMA} \bibinfo{volume}{305}, \bibinfo{pages}{1018}.
\newblock \URLprefix
  \url{http://jama.jamanetwork.com/article.aspx?doi=10.1001/jama.2011.252},
  \DOIprefix\doi{10.1001/jama.2011.252}. \bibinfo{note}{tex.ids:
  rizzoImpairedDrivingMedical2011a}.
\bibitem[{Skurtveit et~al.(2009)Skurtveit, Strøm, Skrivarhaug, Mørland,
  Bramness and Engeland}]{skurtveit_road_2009}
\bibinfo{author}{Skurtveit, S.}, \bibinfo{author}{Strøm, H.},
  \bibinfo{author}{Skrivarhaug, T.}, \bibinfo{author}{Mørland, J.},
  \bibinfo{author}{Bramness, J.G.}, \bibinfo{author}{Engeland, A.},
  \bibinfo{year}{2009}.
\newblock \bibinfo{title}{Road traffic accident risk in patients with diabetes
  mellitus receiving blood glucose-lowering drugs. {Prospective} follow-up
  study}.
\newblock \bibinfo{journal}{Diabetic Medicine} \bibinfo{volume}{26},
  \bibinfo{pages}{404--408}.
\newblock \URLprefix
  \url{http://doi.wiley.com/10.1111/j.1464-5491.2009.02699.x},
  \DOIprefix\doi{10.1111/j.1464-5491.2009.02699.x}.
\bibitem[{Sommerfield et~al.(2004)Sommerfield, Deary and
  Frier}]{sommerfield_acute_2004}
\bibinfo{author}{Sommerfield, A.J.}, \bibinfo{author}{Deary, I.J.},
  \bibinfo{author}{Frier, B.M.}, \bibinfo{year}{2004}.
\newblock \bibinfo{title}{Acute {Hyperglycemia} {Alters} {Mood} {State} and
  {Impairs} {Cognitive} {Performance} in {People} {With} {Type} 2 {Diabetes}}.
\newblock \bibinfo{journal}{Diabetes Care} \bibinfo{volume}{27},
  \bibinfo{pages}{2335--2340}.
\newblock \URLprefix
  \url{https://diabetesjournals.org/care/article/27/10/2335/23192/Acute-Hyperglycemia-Alters-Mood-State-and-Impairs},
  \DOIprefix\doi{10.2337/diacare.27.10.2335}.
\bibitem[{Stork et~al.(2007)Stork, van Haeften and
  Veneman}]{stork_decision_2007}
\bibinfo{author}{Stork, A.D.}, \bibinfo{author}{van Haeften, T.W.},
  \bibinfo{author}{Veneman, T.F.}, \bibinfo{year}{2007}.
\newblock \bibinfo{title}{The {Decision} {Not} to {Drive} {During}
  {Hypoglycemia} in {Patients} {With} {Type} 1 and {Type} 2 {Diabetes}
  {According} to {Hypoglycemia} {Awareness}}.
\newblock \bibinfo{journal}{Diabetes Care} \bibinfo{volume}{30},
  \bibinfo{pages}{2822--2826}.
\newblock \URLprefix
  \url{http://care.diabetesjournals.org/cgi/doi/10.2337/dc06-1544},
  \DOIprefix\doi{10.2337/dc06-1544}.
\bibitem[{Tregear et~al.(2007)Tregear, Rizzo, Tiller, Schoelles, Hegmann,
  Phillips, Greenburg and Anderson}]{tregear_diabetes_2007}
\bibinfo{author}{Tregear, S.J.}, \bibinfo{author}{Rizzo, M.},
  \bibinfo{author}{Tiller, M.}, \bibinfo{author}{Schoelles, K.},
  \bibinfo{author}{Hegmann, K.T.}, \bibinfo{author}{Phillips, B.},
  \bibinfo{author}{Greenburg, M.I.}, \bibinfo{author}{Anderson, G.},
  \bibinfo{year}{2007}.
\newblock \bibinfo{title}{Diabetes and {Motor} {Vehicle} {Crashes}: {A}
  {Systematic} {Evidence}-{Based} {Review} and {Meta}-{Analysis}}, in:
  \bibinfo{booktitle}{Proceedings of the 4th {International} {Driving}
  {Symposium} on {Human} {Factors} in {Driver} {Assessment}, {Training}, and
  {Vehicle}}, \bibinfo{publisher}{University of Iowa},
  \bibinfo{address}{Stevenson, Washington, USA}. pp. \bibinfo{pages}{343--350}.
\newblock \URLprefix
  \url{http://ir.uiowa.edu/drivingassessment/2007/papers/55},
  \DOIprefix\doi{10.17077/drivingassessment.1259}.
\bibitem[{Vardi and Zhang(2000)}]{vardi_multivariate_2000}
\bibinfo{author}{Vardi, Y.}, \bibinfo{author}{Zhang, C.H.},
  \bibinfo{year}{2000}.
\newblock \bibinfo{title}{The multivariate {L1}-median and associated data
  depth}.
\newblock \bibinfo{journal}{Proceedings of the National Academy of Sciences}
  \bibinfo{volume}{97}, \bibinfo{pages}{1423--1426}.
\newblock \URLprefix \url{http://www.pnas.org/cgi/doi/10.1073/pnas.97.4.1423},
  \DOIprefix\doi{10.1073/pnas.97.4.1423}.
\bibitem[{Warren and Frier(2005)}]{warren_hypoglycaemia_2005}
\bibinfo{author}{Warren, R.E.}, \bibinfo{author}{Frier, B.M.},
  \bibinfo{year}{2005}.
\newblock \bibinfo{title}{Hypoglycaemia and cognitive function}.
\newblock \bibinfo{journal}{Diabetes, Obesity and Metabolism}
  \bibinfo{volume}{7}, \bibinfo{pages}{493--503}.
\newblock \URLprefix
  \url{http://doi.wiley.com/10.1111/j.1463-1326.2004.00421.x},
  \DOIprefix\doi{10.1111/j.1463-1326.2004.00421.x}.
\bibitem[{{World Health
  Organization}(2016)}]{world_health_organization_global_2016}
\bibinfo{editor}{{World Health Organization}} (Ed.), \bibinfo{year}{2016}.
\newblock \bibinfo{title}{Global report on diabetes}.
\newblock \bibinfo{publisher}{World Health Organization},
  \bibinfo{address}{Geneva, Switzerland}.
\newblock \bibinfo{note}{OCLC: ocn948336981}.

\end{thebibliography}

\end{document}